\documentclass[twocolumn,preprintnumbers,showpacs,superscriptaddress,
nofootinbib]{
revtex4}
\usepackage{mathrsfs}
\usepackage{amsmath,amssymb,graphicx}
\usepackage {amssymb}
\usepackage{color}
\newcommand{\nc}{\newcommand}
\nc{\bea}{\begin{eqnarray}} \nc{\eea}{\end{eqnarray}}
\nc{\be}{\begin{equation}} \nc{\ee}{\end{equation}}

\newcommand\s{\sigma}

\nc{\ga}{\gamma} \nc{\x}{{\bf x }} \nc{\kk}{{\bf k }} \nc{\f}{{\bf f
}} \nc{\T}{ \theta (s_i (t)- \s) } \nc{\TT}{ \theta (s_i (t_{ r \, i
} )- \s) } \nc{\br}{   (s_i (t)- \s)  } \nc{\fa}{\phi_1}
\nc{\fb}{\phi_2}

\input{epsf.sty}

\begin{document}

\title{Towards Anisotropy-Free and Non-Singular Bounce Cosmology with Scale-invariant Perturbations}

\author{Taotao Qiu}
\email{xsjqiu@gmail.com,qiutt@ntu.edu.tw}
\affiliation{ Leung Center for Cosmology and Particle Astrophysics National Taiwan University, Taipei 106, Taiwan}
\affiliation{ Department of Physics, National Taiwan University, Taipei 10617, Taiwan}

\author{Xian Gao}
\email{xgao@apc.univ-paris7.fr}
\affiliation{Astroparticule et Cosmologie (APC), UMR 7164-CNRS, Universit\'{e} Denis Diderot-Paris 7, 10 rue Alice Domon et L\'{e}onie Duquet, 75205 Paris, France}
\affiliation{Laboratoire de Physique Th\'{e}orique, \'{E}cole Normale Sup\'{e}rieure, 24 rue Lhomond, 75231 Paris, France}
\affiliation{Institut d'Astrophysique de Paris (IAP), UMR 7095-CNRS, Universit\'{e} Pierre et Marie Curie-Paris 6, 98bis Boulevard Arago, 75014 Paris, France}

\author{Emmanuel N. Saridakis}
\email{Emmanuel_Saridakis@baylor.edu}
\affiliation{Physics Division, National Technical University of Athens, 15780 Zografou Campus,  Athens, Greece}
\affiliation{Instituto de F\'{\i}sica, Pontificia Universidad de Cat\'olica de Valpara\'{\i}so, Casilla 4950, Valpara\'{\i}so, Chile}


\begin{abstract}
We investigate non-singular bounce realizations in the framework of
ghost-free generalized Galileon cosmology, which furthermore can be
free of the anisotropy problem.
Considering an Ekpyrotic-like potential we can obtain a total
Equation-of-State (EoS) larger than one  in the contracting phase, which is
necessary for the evolution to be stable against small anisotropic
fluctuations. Since such a large EoS forbids the Galileon field to generate
the desired form of perturbations, we additionally introduce the
curvaton field which can in general produce the observed nearly
scale-invariant spectrum. In particular, we provide approximate analytical
and exact semi-analytical expressions under which the bouncing scenario is
consistent with observations. Finally, the combined Galileon-curvaton
system is free of the Big-Rip after the bounce.
\end{abstract}


\maketitle

\section{Introduction}

Non-singular bouncing cosmology \cite{Novello:2008ra} has gained significant interest in recent studies of the early universe. The main reason for such a research direction is that the most popular paradigm of the early universe, namely inflation, still suffers from the ``Big-Bang singularity'' problem, which however can be naturally avoided in non-singular bouncing or cyclic cosmologies. Additionally, these paradigms can also solve the horizon, flatness and monopole problems, and make compatible observational predictions such as nearly scale-invariant power spectrum and moderate non-Gaussianities \cite{Cai:2007zv,Cai:2008ed,Qiu:2010ch}. Therefore, they are recently considered as good alternatives to inflation.

In order to realize a successful bounce several requirements must be
fulfilled. First of all, the basic condition is to have the Hubble
parameter change its sign from negative to positive at the bounce, which
implies that during the bouncing phase the Null Energy Condition (NEC) must
be violated, with the total EoS of the universe going below $-1$
\cite{Cai:2007qw,Cai:2009zp}. The NEC violation in the context of General
Relativity is nontrivial \cite{Nojiri:2013ru}, usually leading to ghost
degree(s) of freedom \cite{Carroll:2003st,Cline:2003gs}, which would demand
either ghost-elimination mechanisms or an extended analysis to a modified
gravity context \cite{Capozziello:2011et,Biswas:2005qr}.

Apart from the above basic condition, in order for a bounce to be a
successful alternative to inflation it should also solve the other Big-Bang
problems, and moreover it should produce a nearly scale-invariant power
spectrum as required by observations \cite{Bennett:2012fp}. These
impose more stringent constraints on the bounce evolution, especially in
the contracting phase. For instance, the horizon problem can be solved if
the quantum fluctuations in the far past lie deep inside the horizon, while
they should exit the horizon in the contracting phase in order to generate
perturbations compatible with observations, provided that inflation is
absent in bouncing scenario. This requires a total EoS satisfying $w>-1/3$
in the contracting phase \cite{Piao:2004jg,Qiu:2012ia}. However, the
scale-invariance of the perturbations is even harder to be achieved. In
particular, as it was initially shown in \cite{Finelli:2001sr}, if
the perturbations generated in the contracting phase are purely adiabatic,
the EoS of the contracting universe should satisfy $w\approx0$ in order to
produce the desired spectrum.

However, although bounce models with total EoS $w\approx0$ before the
bounce, namely the ``matter bounce'', could lead to nearly scale-invariant
power spectrum, they generally suffer from the ``anisotropy problem''
\cite{Kunze:1999xp} in the contracting phase. In
particular, in 4D General Relativity  a tiny amount of anisotropic
fluctuation from the simple isotropic Friedmann-Robertson-Walker (FRW)
geometry in the contracting phase, would increase as $a^{-6}$, where $a$ is
the scale factor. Thus it would finally dominate over
matter-like background, leading to a Big-Crunch singularity with complete
anisotropy instead of a bounce, unless one impose a strong fine-tuning of
the model parameters and the initial amount of anisotropy in order to
obtain a bounce before the domination of the anisotropic term. In that
sense the ``matter bounce" scenario is not stable against cosmological
anisotropy (for its similar problem in the presence of radiation see
\cite{Karouby:2010wt}). For this reason, we must construct scenarios with
total EoS larger than $1$ in order to prevent the dominance of anisotropy.
However, as we mentioned above, a different EoS may not be able to provide
the scale-invariant power spectrum, if we insist on applying the simple
adiabatic mechanism of generating primordial perturbations\footnote{We
would like to mention that here by ``simple adiabatic mechanism'' we mean
that the perturbations generated are purely adiabatic, and the EoS remains
constant. However, the term ``adiabatic mechanism'' which was first
proposed in \cite{Khoury:2009my} in ``Ekpyrotic'' scenarios
\cite{Khoury:2001wf}, refers to the mechanism that generates adiabatic
perturbations via varying EoS.}. Therefore, we should resort to alternative
mechanisms, such as adiabatic, entropy and conformal ones
\cite{Khoury:2009my,Finelli:2002we,Lehners:2007ac,
Buchbinder:2007ad,Lehners:2007wc, Hinterbichler:2011qk}
\footnote{Note that the stability of isotropic solutions in anisotropic
perturbations has been studied in \cite{Aref'eva:2009vf}.}.

In the present work we investigate the bounce realization in the framework
of recently proposed generalized Galileon cosmology
\cite{Nicolis:2008in,Deffayet:2009mn} (see also
\cite{Silva:2009km,DeFelice:2011bh} for various
developments). Due to the delicate design of the Lagrangian form such
a theory, which contains higher-order derivatives, can keep its equation of
motion up to second-order and thus is free of ghosts (this was pioneered by
the work by Horndeski \cite{Horndeski:1974wa}), but it can
indeed provide extra degree(s) of freedom in order to violate NEC.
Recently, in  \cite{Qiu:2011cy}, the first ghost-free bounce model
based on Galileon cosmology was constructed by one of the present
authors and collaborators (see also \cite{Easson:2011zy}), and hence in
this article we will consummate this class of models by addressing the
problems mentioned above. Note that alternative scenarios addressing the
anisotropy problem in Galileon bouncing cosmologies have been presented
in \cite{Cai:2012va,Cai:2013vm}, of which before contracting with $w>1$, the
universe
can be dominated by cold matter \cite{Lin:2010pf}, where scale-invariant
perturbations could be generated.

 First of all, by introducing an Ekpyrotic-like
negative potential  we can easily obtain a very large EoS in the
contracting phase, thus the anisotropy problem will be eliminated. However,
as mentioned above, a large EoS forbids the Galileon field to generate the
desired form of perturbations, thus as a next step  we additionally
introduce the curvaton field which is suitably coupled to the Galileon
field, such that the nearly scale-invariant spectrum can be produced.
Finally, we perform a complete analysis of the behavior around the bounce
point of the full Galileon-curvaton system, making use of the ``inverse"
reconstruction procedure \cite{Cai:2009in}, showing that with a proper
choice of the Lagrangian functional forms a non-singular bounce
can be reconstructed, which can connect smoothly to the matter-domination
era and moreover alleviate the Big-Rip singularity which
appears in \cite{Qiu:2011cy}.

The plan of the work is the following: In section \ref{anisotropyproblem}
we briefly review the anisotropy problem. In section \ref{Galileonbounce}
we present the bouncing background evolution before,
during and after the bounce, and we show   that the
perturbations are stable and free of ghosts. In section
\ref{curvatonmechanism} we analyze the curvaton mechanism that produces
nearly scale-invariant perturbations. In section \ref{Reconstructing} we
perform a semi-analytical procedure in order to reconstruct an exact
bouncing solution that is not followed by a Big-Rip. Finally, in section
\ref{Conclusions} we summarize and we discuss the obtained results.
Throughout the manuscript we use the  $(-,+,+,+)$ metric signature,
 and units in which $M_{Pl}=1/\sqrt{8\pi G}=1$.

\section{The anisotropy problem}
\label{anisotropyproblem}

The anisotropy problem is a notorious problem that generally exists in
bouncing models with $w<1$ in contracting phase
\cite{Kunze:1999xp}. In General Relativity, if we
allow the existence of a non-zero anisotropy at the beginning of the
contraction  it will evolve scaling as $a^{-6}(t)$. In order to demonstrate
this more transparently, without loss of generality we consider as an
example the simple anisotropic Bianchi-IX metric \cite{Misner:1974qy}:
\be
ds^2=-dt^2+a^2(t)\sum_{i=1}^3e^{2\beta_i(t)}d{x^i}^2~,
\ee
with $\beta_1(t)+\beta_2(t)+\beta_3(t)=0$. The Friedmann Equation writes as:
\be\label{friedmannani}
3H^2=\rho_u+\frac{1}{2}\left(\sum_{i=1}^3\dot\beta_i^2\right)~,
\ee
where $H=\dot a/a$ the Hubble parameter  and with $\rho_u$ incorporating
all the fluids in the universe. The $\beta_i$'s satisfy the equations
\be
\ddot\beta_i+3H\dot\beta_i=0~,
\ee
which provide the solutions   $\beta_i\propto a^{-3}(t)$. Since the
second part in the right hand side of equation (\ref{friedmannani}) can be
considered as an effective anisotropy term $\chi^2$, we conclude that
\be
\chi^2\equiv\frac{1}{2}\left(\sum_{i=1}^3\dot\beta_i^2\right)\propto
a^{-6}(t)~,
\ee
and thus the anisotropy term corresponds to an effective energy density
with EoS $w=1$. Although in an expanding universe this term is always
sub-dominant and thus isotropization can be achieved, in a contracting
case, as long as it is initially non-zero (even arbitrarily small), the
anisotropy will grow fast and become dominant over all species with EoS
less than 1, leading finally to a collapsing anisotropic universe. For this
reason, in order to avoid the domination of a possible anisotropic
fluctuation, one has to realize a contracting background that evolves even
faster, which requires an EoS larger than unity in the contracting phase
\footnote{For some Grand Unification Theories where anisotropic stresses
and collisionless particles are taken into account, there may still be
anisotropy problems, see \cite{Barrow:2010rx} for more details. However,
it is not the case that we're currently considering. We thank John Barrow
for pointing it out to us.}.

\section{The Galileon bounce}
\label{Galileonbounce}

In the previous section we briefly showed that in order to realize a bounce
we need an effective EoS $w>1$ in the contracting phase, in order to avoid
the domination of an anisotropic fluctuation.
In this section we formulate the bounce realization in generalized Galileon cosmology.

In the generalized Galileon scenario, where the coefficients of the various
action-terms are considered as functions of the scalar field, the
corresponding action can be written as \cite{Deffayet:2009mn}:
\begin{equation}
{\cal L}=\sum_{i=2}^{5}{\cal L}_{i}\,,\label{Ltotal}
\end{equation}
where
\begin{eqnarray}
&&{\cal L}_{2} = K(\phi,X)\nonumber\\
&&{\cal L}_{3} = -G_{3}(\phi,X)\Box\phi\nonumber\\
&&{\cal L}_{4} = G_{4}(\phi,X)\,
R+G_{4,X}\,[(\Box\phi)^{2}-(\nabla_{\mu}\nabla_{\nu}\phi)\,(\nabla^{\mu}
\nabla^{\nu}\phi)]\nonumber\\
&&{\cal L}_{5} = G_{5}(\phi,X)\,
G_{\mu\nu}\,(\nabla^{\mu}\nabla^{\nu}\phi)\nonumber\\
&&\ \ \ \ \ \ \  -\frac{1}{6}\,
G_{5,X}\left[(\Box\phi)^{3}
 -3(\Box\phi)\,(\nabla_{\mu}\nabla_{\nu}\phi)\,
(\nabla^{\mu}\nabla^{\nu}\phi)\right.
\nonumber\\
&&\left.\ \ \ \ \ \ \ \ \ \ \ \ \ \ \ \ \ \ \ \
+2(\nabla^{\mu}\nabla_{\alpha}\phi)\,(\nabla^
{\alpha}\nabla_{\beta}\phi)\,(\nabla^{\beta}\nabla_{\mu}\phi)\right]\,
.\label {
eachlag5}
\end{eqnarray}
In this action the functions $K$ and $G_{i}$ ($i=3,4,5$) depend on the
scalar field $\phi$ and its kinetic energy
$X\equiv-\frac{1}{2}\nabla_\mu\phi\nabla^\mu\phi$, while $R$ is the Ricci
scalar and $G_{\mu\nu}$ is the Einstein tensor. Moreover, $G_{i,X}$ and
$G_{i,\phi}$ ($i=3,4,5$) denote the partial derivatives of $G_{i}$ with
respect to $X$ and $\phi$, ($G_{i,X}\equiv\partial G_{i}/\partial X$ and
$G_{i,\phi}\equiv\partial
G_{i}/\partial\phi$), and the box operator is constructed from covariant derivatives: $\Box\phi\equiv g^{\mu\nu} \nabla_\mu \nabla_\nu\phi$. In the following we focus on the case
\begin{eqnarray}
\label{model}
&&K(\phi,X)=X-V(\phi),~G_3(\phi,X)=g(\phi)X,\nonumber\\
&&G_4(\phi,X)=
\frac{1}{2}, ~G_5(\phi,X)=0.
\end{eqnarray}
Therefore, the action that we are going to use reads:
\bea
\label{action}
&&{\cal
S}=\int
d^4x\sqrt{-g}\left[\frac{1}{2}R-\frac{1}{2}
\nabla_\mu\phi\nabla^\mu\phi-V(\phi)\right.\ \  \ \   \   \ \   \ \   \   \
\
\nonumber\\
&&\left.\ \ \ \  \ \   \   \ \    \ \   \   \ \  \ \   \   \ \   \
\ \     +\frac{g}{2}
\nabla_\mu\phi\nabla^\mu\phi
\Box\phi \right].
\eea

We now proceed to a detailed investigation of the above scenario. Firstly,
in the following subsection we provide approximate analytical solutions at
the far past before the bounce, around the bouncing regime, and after the
bounce, at the background level. Then in the next subsection, we analyze
the perturbation behavior.

\subsection{Background evolution: analytical results}
\label{galileonbackground}

In the following we impose a flat Friedmann-Robertson-Walker (FRW) background metric of the form $ds^{2}=-N^{2}(t)dt^{2}+a^{2}(t)d{\bf{x}}^{2}$, where $t$ is the cosmic time, $x^i$ are the comoving spatial coordinates, $N(t)$ is the lapse function, and $a(t)$ is the scale factor. Varying the action (\ref{action}) with
respect to $N(t)$ and $a(t)$ respectively, and setting $N=1$, we obtain the Friedmann equations
\be
\label{friedmann}
 H^2=\frac{1}{3}\rho~,~~~\dot{H}=-\frac{1}{2}(\rho+P)~.
\ee
Additionally, we have defined the effective energy density and pressure as:
\bea
\label{rho}
 \rho &=& \frac{1}{2}\dot\phi^2+V(\phi)+3gH\dot\phi^3~,\\
\label{pressure} P &=&
\frac{1}{2}\dot\phi^2-V(\phi)-g\dot\phi^2\ddot\phi~,
\eea
and thus the total EoS of the universe is just
\bea
\label{wtot}
w\equiv\frac{P}{\rho}=\frac{\dot\phi^2-2V(\phi)-2g\dot\phi^2\ddot\phi}{\dot\phi^2+2V(\phi)+6gH\dot\phi^3}~.
\eea
Finally, variation of (\ref{action}) with respect to  the Galileon field provides its evolution equation:
\be
\label{eom}
\mathcal{D}\ddot{\phi}+\Gamma\dot{\phi}+V_{\phi}=0~,
\ee
where
\bea
\label{mathcalD}
\mathcal{D}&=&1+6gH\dot{\phi}+\frac{3}{2}g^{2}\dot{\phi}^{4}~, \\
\Gamma&=&\frac{3}{2}\left(1+3gH\dot{\phi}\right)\left(2H-g\dot{\phi}^{3}
\right)~.
\eea

In the following we will suitably choose the potential $V(\phi)$ in
order to obtain a very large positive equation of state in relation
(\ref{wtot}), namely $w>1$, so that our model will not suffer from the
anisotropy problem in the contracting phase. Meanwhile, when the Galileon
term $G\Box\phi$ becomes more and more important, the EoS becomes negative
and eventually triggers the bounce.

As a specific example, we choose the potential to be
\be
V(\phi)=-V_0e^{c\phi},
\label{potential}
\ee
with $V_0,c>0$, namely a negative exponential potential, which is the usual
one in Ekpyrotic scenarios \cite{Khoury:2001wf}. Within this choice, when
the nonlinear kinetic term is relatively small we obtain $\rho<P$ and thus
$w>1$. One could ask whether the negative potential would lead to a
negative energy density, however the more negative the potential is, the
steeper it is, and thus it gives rise to larger kinetic term, which can
compensate the potential negativity (this will be verified later on).

Let us proceed to a qualitative investigation of the dynamics of such
scenario. For simplicity we consider that $\phi$ increases from negative to
positive during the evolution, therefore at the beginning when $\phi$
starts at a large negative value the potential (\ref{potential}) is in its
``slow-varying'' region, and the field  moves slowly. In this region the
nonlinear term $G\Box\phi$ will have a small contribution to the action.
Similarly to the Ekpyrotic models, the universe will contract with a very
large positive EoS, but as time passes $\phi$ moves towards positive
values and its velocity increases, the effect of the $G\Box\phi$ term is
enhanced, and it can trigger the bounce. However, after the bounce and due
to the large slope of the potential, the field kinetic energy could
increase to unacceptably large values which could spoil the validity of the
effective theory. Therefore, we should need some mechanisms to slow the
field motion down, and this will be demonstrated in detail in the next
sections.

In the following we proceed to the quantitative investigation of the
scenario, extracting approximate analytical solutions for the background
evolution in the far past before the bounce, around the bouncing point,
and in the far future after the bounce.

\subsubsection{Solution far before the bounce}

Far before the bounce, as we assumed, $\phi$ begins with a large negative value and moves slowly, while the nonlinear term of $\dot\phi$ is negligible. The energy density and pressure reduces to
\be
\rho\simeq\frac{1}{2}\dot\phi^2+V(\phi)~,~P\simeq\frac{1}{2}\dot\phi^2-V(\phi)~,
\ee
similarly to a single scalar field with an Ekpyrotic potential. Thus, we can choose the model parameters and the initial conditions of $\phi$ in order to acquire scaling solutions, namely to obtain a scale factor evolving as
\be
\label{scalea}
 a(t)\sim (t_\ast-t)^p\sim|\eta_\ast-\eta|^\frac{p}{1-p}
~,~\ \ p\equiv\frac{2}{3(1+w)},~
\ee
where the subscript `$\ast$' denotes the time where the non-linear term in
equations (\ref{rho})-(\ref{eom}) becomes important. Note that since we are
considering the region where $t<t_\ast$, we have $t_\ast-t>0$.
For completeness we have also expressed the above solution using  the
conformal time $\eta$, related to the cosmic time $t$ through
$dt=ad\eta$. Similarly,
the field $\phi(t)$ scales as
\be
\label{ansatz}
\phi(t)\simeq-\frac{2}{c}\ln(t_\ast-t)~,~p=\frac{2}{c^2}~.
\ee
Moreover, taking the derivatives of the above expressions we find
\bea
\label{dotphi0}
&&\dot\phi(t)\simeq\frac{2}{c(t_\ast-t)},\\
&&H(t)\simeq\frac{p}{
t-t_\ast } ~,
\label{Hsol0}
\eea
and therefore the energy density from expression (\ref{rho}) becomes
\bea
\label{energybg}
\rho(t)_{t<t_\ast}=3H^2\propto (t-t_\ast)^{-2}
\eea
Finally, using relations (\ref{scalea}) and (\ref{ansatz}), for the total
EoS using (\ref{wtot}) we obtain
\bea
\label{wtot1}
w\simeq\frac{1}{3}c^2-1=const.
\eea

\subsubsection{Solution around the bounce}

Around the bouncing point $t_B$ the nonlinear term becomes important. From the Friedmann equation (\ref{friedmann}) we can express $H$ as:
\be
\label{solH} H=\frac{1}{2}g\dot\phi^3\pm\frac{1}{6}\sqrt{
9g^2\dot\phi^6+6(\dot\phi^2-2V_0e^{c\phi})}~,
\ee
in which the nonlinear term becomes important and the last terms in
(\ref{rho}) and (\ref{pressure}) can no longer be neglected. Without loss
of generality we can keep the minus-sign branch in order to obtain a
positive $\dot\phi(t)$.\footnote{Note that in general the Hubble parameter
could transit from the minus-sign branch to the plus-sign branch either
before (for $\dot\phi_{initial}<0$) or after the bounce (for
$\dot\phi_{initial}>0$). Thus, above we assume that if such transition
exist it will take place after the bounce. However, as we will shortly see
below, at late times and under the curvaton  backreaction,
expression (\ref{solH}) is not exactly valid any more. Hence, we do not
examine in detail the relation between the two branches.} Therefore, when
the universe goes from the contracting ($H<0$) to the expanding phase
($H>0$), one has $\dot\phi>\sqrt{2V_0}e^{c\phi/2}$ before and
$\dot\phi<\sqrt{2V_0}e^{c\phi/2}$ after the bounce. It is therefore natural
to consider the solution at the bounce region to be
\be\label{phibounce}
\dot\phi=\sqrt{2V_0}(\alpha t+\beta)e^{\frac{c\phi}{2}}~,
\ee
with $\alpha$ and $\beta$ being two parameters satisfying the conditions
\be
\alpha<0~,~\alpha t_B+\beta=1~.
\ee
Note that substitution of (\ref{phibounce}) into the $\phi$-equation of
motion (\ref{eom}), provides the necessary value for
the coefficient
$\alpha$ in order to have self-consistency.

Integrating equation (\ref{phibounce}) leads to
\be
e^{-\frac{c\phi_0}{2}}-e^{-\frac{c\phi}{2}}=\sqrt{2V_0}\Big[\frac{\alpha}{2
}(t_0^2-t^2)+\beta(t_0-t)\Big]~,
\ee
with $t_0$ a boundary value for $t$, and $\phi_0$ the corresponding value of $\phi$. The above formula will be simplified if we set $t_0$ in the far future when $\phi_0$ becomes very large, and thus $e^{-\frac{c\phi_0}{2}}\approx 0$. In this case the expression for $\phi(t)$ becomes
\be
\phi=-\frac{2}{c}\ln\Big\{-\sqrt{2V_0}(t_0-t)\Big[\frac{\alpha}{2}
(t-t_0)+\alpha t_0+\beta\Big]\Big\}~,
\label{solfull}
\ee
which leads to
\bea
\label{solfull1}
\dot\phi&=& \frac{4(\alpha t+\beta)}{ c(t_0-t)[\alpha(t-t_0)+2(\alpha
t_0+\beta)]}  ~,\\
\ddot\phi&=&\frac{2}{c}\left\{\frac{1}{(t_0-t)^2}+\frac{\alpha^2}{[
\alpha(t+t_0)+2\beta]^2}\right\}~.
\label{solfull2}
\eea

It would be useful if we could approximate the above expressions around the bouncing point $t_B$, namely $t\rightarrow t_B=(1-\beta)/\alpha$. In this case, and neglecting the constant terms in order to extract the pure scaling behavior, we obtain
\bea
&&\phi\simeq
 \frac{4 \alpha(t-t_B)}{c\left[(\beta +\alpha
t_0)^2-1\right]},\nonumber\\
&&\dot\phi\simeq\frac{4\alpha^2\left[(\beta+\alpha
t_0)^2+1\right](t-t_B)}{c\left[ (\beta+\alpha
t_0)^2-1\right]^2},
\nonumber\\
&&\ddot\phi\simeq\frac{8\alpha^3\left[3(\beta+\alpha
t_0)^2+1\right](t-t_B)}{c  (\beta+\alpha t_0-1)^3(\beta+\alpha t_0+1)^3},
\label{phiapprox0}
\eea
that is $\phi$ exhibits a linear behavior around $t_B$. Following the
same way, one could also insert these expressions into equation
(\ref{solH}) and (\ref{wtot}) to straightforwardly obtain the approximate
 solutions for $H(t)$ and $w(t)$, respectively.

\subsubsection{Solution after the bounce}

Far after the bounce, when $t$ approaches $t_0$, the solutions are still
given by   (\ref{solfull})-(\ref{solfull2}). Thus, approximating
them at $t\rightarrow t_0$,  and neglecting the constant terms in order to
extract the pure scaling behavior, we acquire
\bea
\label{phiapprox}
\phi&\simeq&-\frac{2}{c}\ln(t_0-t)~,
\\
\dot\phi&\simeq&
\frac{2}{c}\frac{1}{t_0-t}~,
\\
\ddot\phi& \simeq&\frac{2}{c}\frac{1}{(t_0-t)^2}~.
\eea

In this case, inserting these expressions into (\ref{solH}), we can extract a
simple approximate expression for $H(t)$ too, namely
\be
\label{scaleH}
H\simeq \frac{1}{(t_0-t)^3}~,
\ee
which leads to a Big-Rip singularity when $t$ approaches $t_0$.
This can be easily explained since when the nonlinear terms become very
important the last terms in (\ref{rho}) and (\ref{pressure}) become
dominant. Thus, when the energy density is dominated by the $3gH\dot\phi^3$
term, namely $3H^2=\rho\sim 3gH\dot\phi^3$, we straightforwardly find that
the Hubble parameter $H$ will be proportional to $\dot\phi^3$, and
(\ref{scaleH}) is verified. Finally, the total EoS can also be obtained
by inserting these expressions into (\ref{wtot}).

\subsubsection{Numerical verification}
\begin{figure}[ht]
\begin{center}
\includegraphics[scale=0.31]{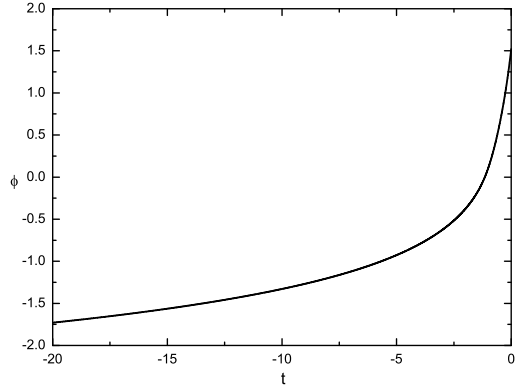}
\caption{{\it{The evolution of the Galileon field  $\phi$ with respect to
$t$. We choose
the initial conditions to be $\phi_i=-\sqrt{3}(\ln20)/3$ and
$\dot\phi_i=\sqrt{3}/60$, and the parameters to be $c=2\sqrt{3}$, $g=1$,
and $V_0=1/12$, respectively.}}} \label{phi}
\end{center}
\end{figure}
\begin{figure}[ht]
\begin{center}
\includegraphics[scale=0.31]{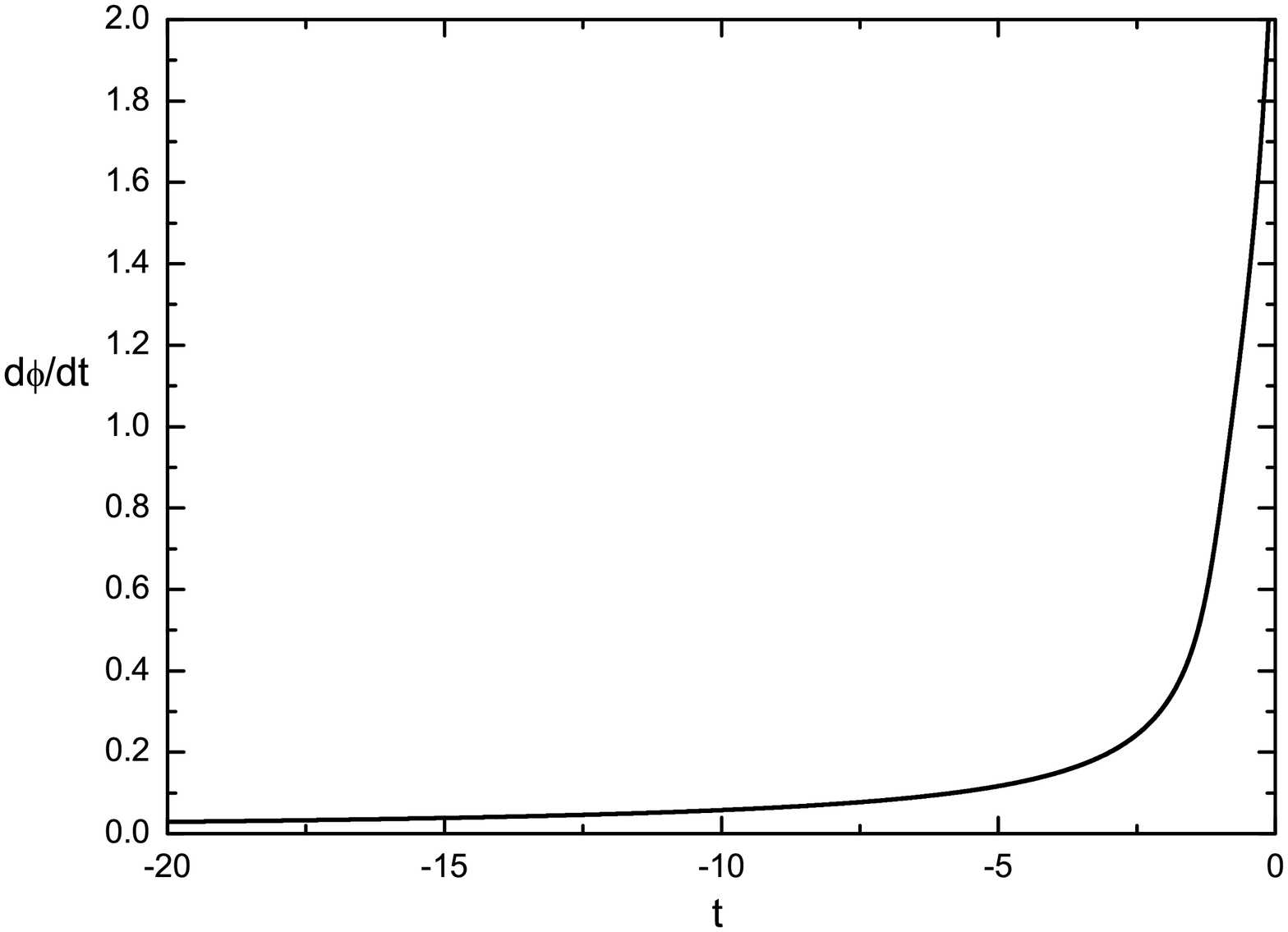}
\caption{{\it{The evolution of the speed of the Galileon field $\dot\phi$
with respect
to $t$. We choose the initial conditions to be $\phi_i=-\sqrt{3}(\ln20)/3$
and $\dot\phi_i=\sqrt{3}/60$, and the parameters to be $c=2\sqrt{3}$,
$g=1$, and $V_0=1/12$, respectively.}}} \label{dotphi}
\end{center}
\end{figure}
\begin{figure}[ht]
\begin{center}
\includegraphics[scale=0.31]{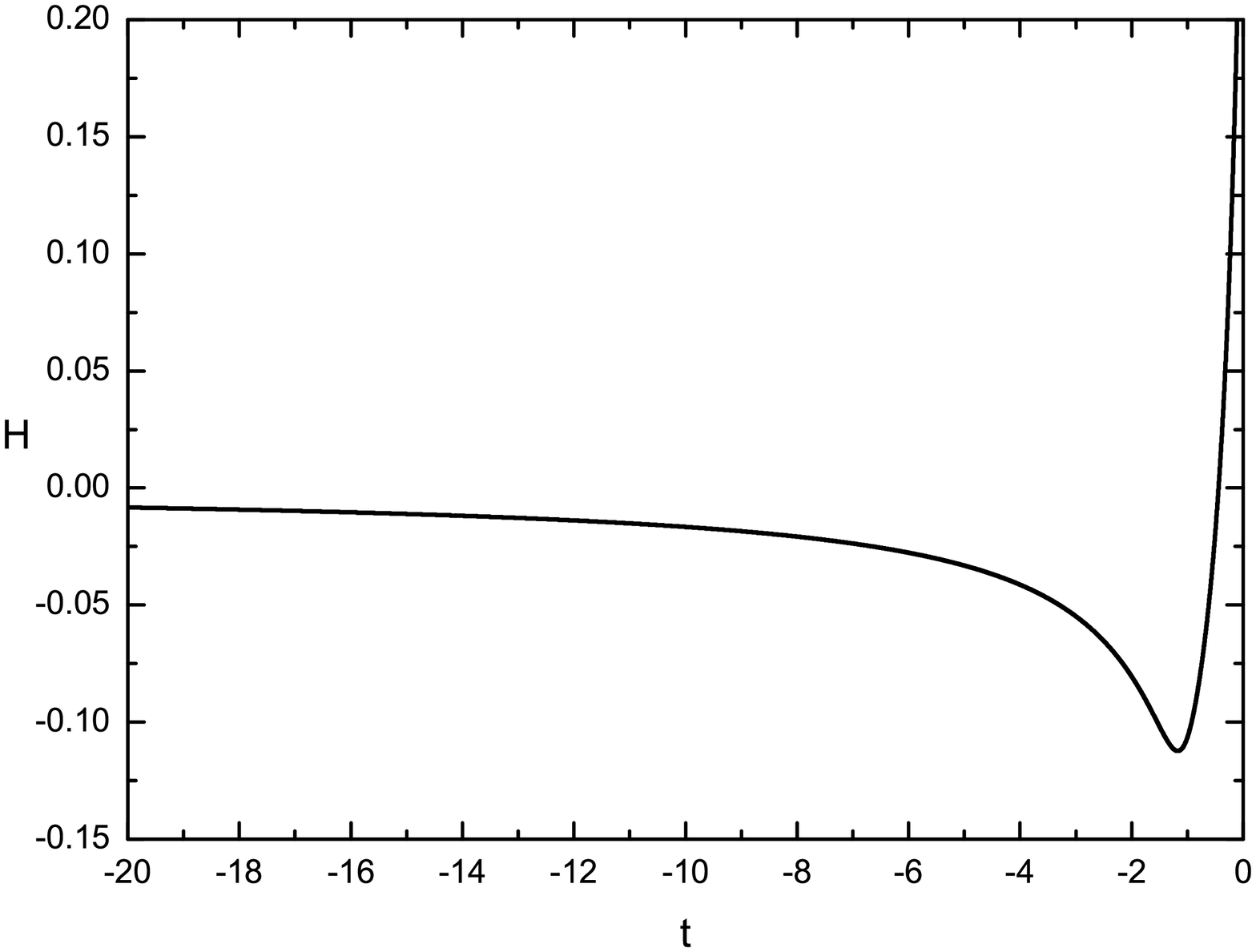}
\caption{{\it{The evolution of the Hubble parameter $H$ with respect to
$t$. We
choose the initial conditions to be $\phi_i=-\sqrt{3}(\ln20)/3$ and
$\dot\phi_i=\sqrt{3}/60$, and the parameters to be $c=2\sqrt{3}$, $g=1$,
and $V_0=1/12$, respectively.}}} \label{hubble}
\end{center}
\end{figure}
\begin{figure}[ht]
\begin{center}
\includegraphics[scale=0.31]{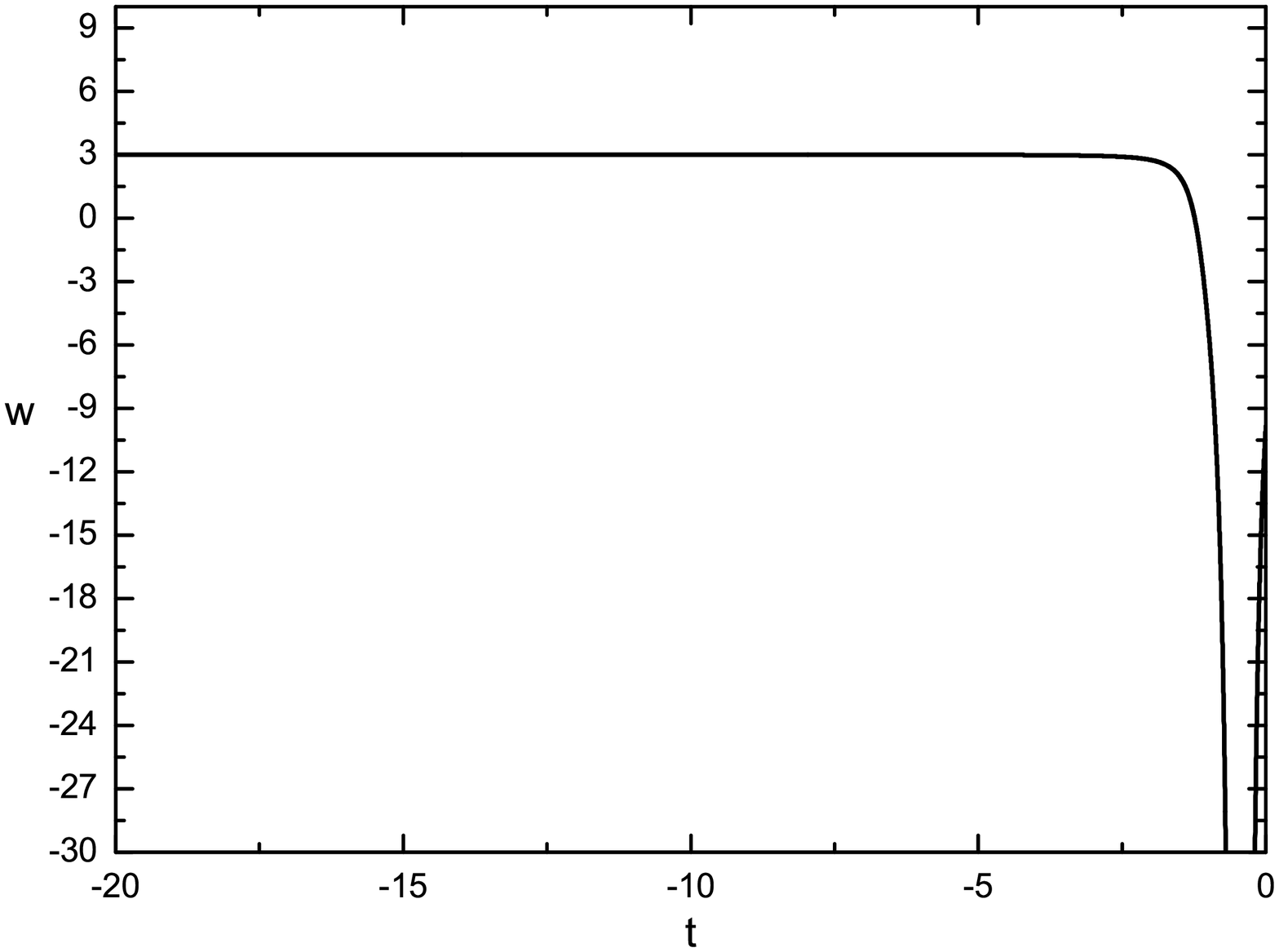}
\caption{{\it{The evolution of the EoS $w$ with respect to $t$. We choose
the
initial conditions to be $\phi_i=-\sqrt{3}(\ln20)/3$ and
$\dot\phi_i=\sqrt{3}/60$, and the parameters to be $c=2\sqrt{3}$, $g=1$,
and $V_0=1/12$, respectively.}}} \label{eos}
\end{center}
\end{figure}
\begin{figure}[ht]
\begin{center}
\includegraphics[scale=0.31]{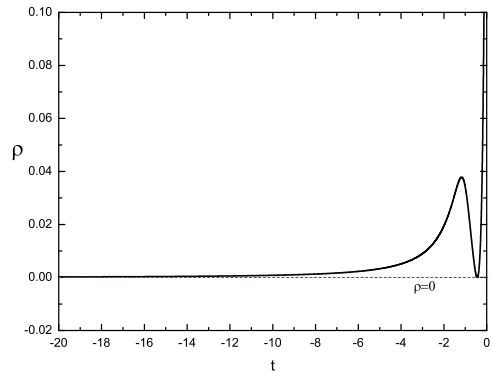}
\caption{{\it{The evolution of the total energy
density $\rho$ of the Galileon field, with respect to $t$. We choose
the
initial conditions to be $\phi_i=-\sqrt{3}(\ln20)/3$ and
$\dot\phi_i=\sqrt{3}/60$, and the parameters to be $c=2\sqrt{3}$, $g=1$,
and $V_0=1/12$, respectively.}}} \label{rhoplot}
\end{center}
\end{figure}
We close the background investigation by performing an exact numerical
elaboration in order to verify the above approximate expressions in the
various regimes. In particular, we numerically solve equations
(\ref{friedmann}) and (\ref{eom}), imposing  (\ref{ansatz}) as our
initial conditions. In Figures \ref{phi}, \ref{dotphi}, \ref{hubble} and
\ref{eos}, we respectively present $\phi(t)$, $\dot\phi(t)$, the Hubble
parameter $H(t)$ as well as the EoS $w(t)$. As we observe, both $\phi(t)$
and $\dot\phi(t)$ are monotonically increasing. Moreover, before the bounce
the universe contracts with a scaling solution with EoS $w$ being constant
and larger than unity (in this specific example $w=3$), and thus the
scenario is free from the anisotropy problem discussed in section
\ref{anisotropyproblem}.

As time passes the nonlinear term becomes important and triggers the
bounce, which forces $H(t)$ to change from negative to positive. Note that
the numerical results confirm that $\dot\phi(t)$ has a positive value
during the bouncing period, and thus it justifies our choice of the minus
sign in (\ref{solH}). After the bounce, the nonlinear kinetic term makes
the scalar-field energy density increasing, leading the universe to a
Big-Rip. One can also see that the total EoS $w(t)$ indeed indicates the
bounce followed by the Big-Rip, verifying the analytical results that has
been obtained in preceding paragraphs.

Finally, in Fig. \ref{rhoplot}, we present the evolution of the total energy
density $\rho$ of the Galileon field, calculated through (\ref{rho}). As
we observe $\rho$ is always positive, despite the use of a negative
potential, due to the increase of the kinetic energy, and it becomes zero
only at the bounce point as expected. Therefore, in the present work we do
not need mechanisms that could transit the universe to positive potential
energy \cite{Felder:2002jk}, however it would be desirable to consider a
mechanism that could smooth the increase of the Galileon kinetic
energy, by either considering a bound in the potential, or couple
$\phi$ to other matter fields such is radiation, which could lead to
energy transfer
away from it (a procedure that could lead to the universe preheating too).
These mechanisms lie beyond the scope of the present work and are left for a
future investigation.

\subsection{Perturbations}

In the previous subsection we analyzed the background evolution of the
Galileon  bounce. Thus, we can now proceed to the investigation of the
perturbations,  focusing on their stabilities. It proves convenient to
foliate the FRW metric in an Arnowitt-Deser-Misner (ADM) form
\cite{Arnowitt:1962hi}:
\be
ds^{2}=-N^{2}dt^{2}+h_{ij}(dx^{i}+N^{i}dt)(dx^{j}+N^{j}dt)~,
\ee
where $N$ is the lapse function, $N^i$ is the shift vector, and $h_{ij}$ is the induced 3-metric. One can then perturb these functions as:
\be
N=1+A~,~N_i=\partial_i\psi~,~h_{ij}=a^2(t)e^{2\zeta}\delta_{ij}~,
\ee
where $A$, $\psi$ and $\zeta$ are the scalar metric perturbations. As usual it is useful to define the (gauge-invariant) comoving curvature perturbation through
\be
\mathcal{R}\equiv\zeta+\frac{H}{\dot\phi}\delta\phi~,
\label{curvpert}
\ee
and hence in the uniform $\phi$ gauge we acquire $\delta\phi=0$ and
$\zeta=\mathcal{R}$.

Under the above perturbation scheme, the action (\ref{action}) perturbed up
to second order becomes:
\be
\label{pertaction}
\mathcal{S}^{(2)}=\int d\eta d^3xa^2\frac{Q_{\cal R}}{c_{s}^2}\Bigl[{\cal
R}^{\prime 2}-c_{s}^2(\partial{\cal R})^2\Bigr]~,
\ee
where $\eta$ is the conformal time. In the above expression we have
introduced the sound-speed squared $c_s^2$, and the quantity $Q_{\cal R}$
related to instabilities, which in our specific scenario read as
\bea
c_s^2&=&Q_{\cal
R}^{-1}[1+2g(\ddot\phi+2H\dot\phi)-\frac{1}{2}g^2\dot\phi^4]~,
\\
Q_{\cal
R}&=&1+6gH\dot\phi+\frac{3}{2}g^2\dot\phi^4~.
\eea
\begin{figure}[ht]
\begin{center}
\includegraphics[scale=0.3]{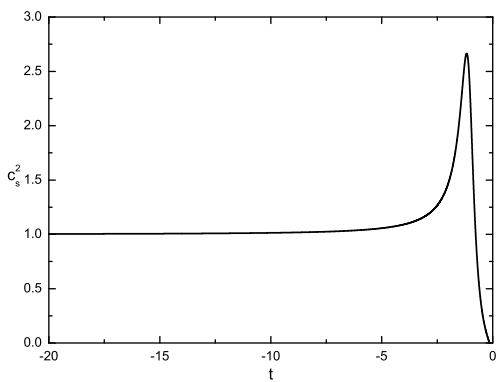}
\caption{{\it{The evolution of the squared speed of sound  $c_s^2$ with
respect
to $t$. We choose the initial conditions to be $\phi_i=-\sqrt{3}(\ln20)/3$
and $\dot\phi_i=\sqrt{3}/60$, and the parameters to be $c=2\sqrt{3}$,
$g=1$, and $V_0=1/12$, respectively.}}}
 \label{cs2}
\end{center}
\end{figure}
Note that according to the definition of $\mathcal{D}$ in (\ref{mathcalD})
we obtain $2Q_{\cal R}=\mathcal{D}$.

\begin{figure}[ht]
\begin{center}
\includegraphics[scale=0.3]{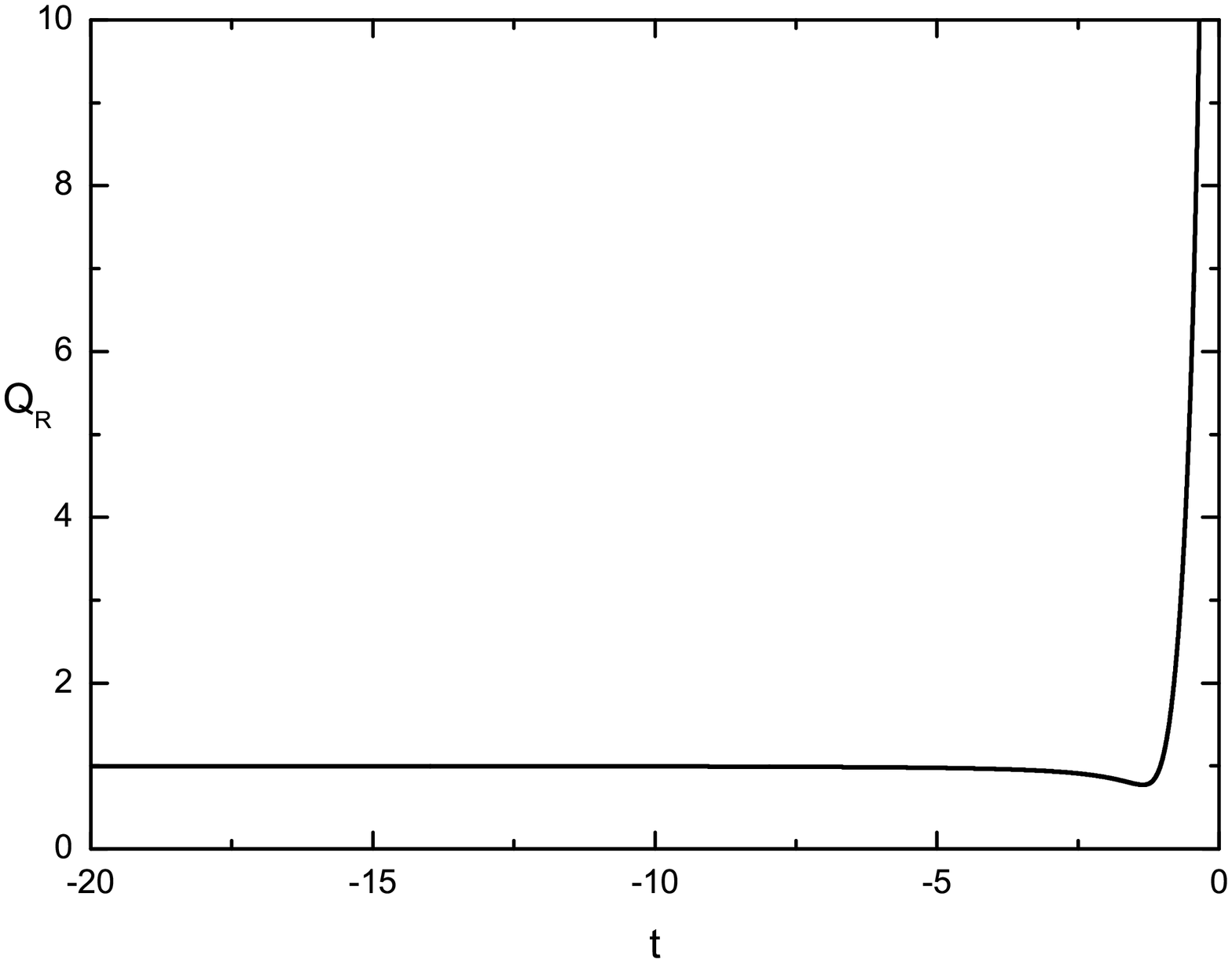}
\caption{{\it{The evolution of the instability-related quantity $Q_{\cal
R}$ with respect to $t$. We choose the
initial conditions to be $\phi_i=-\sqrt{3}(\ln20)/3$ and
$\dot\phi_i=\sqrt{3}/60$, and the parameters to be $c=2\sqrt{3}$, $g=1$,
and $V_0=1/12$, respectively.}}}
 \label{QR}
\end{center}
\end{figure}

The perturbative action (\ref{pertaction}) could in principle lead to ghosts
and gradient instabilities, which would be catastrophic since it is this
action that can be written as a canonical form and then be quantized. From
its form one can see that the avoidance of ghosts requires the factor in
front of the kinetic term of the perturbation variable  ${\cal R}$  to be
positive, namely   $Q_R/c_s^2>0$, while the absence of gradient
instabilities  requires  $c_s^2\geq0$ \cite{DeFelice:2011bh}, which means the
ratio of the factors of spatial and time derivatives must be positive.
In order to show that this is the case for
the exact behavior too, in Figures \ref{cs2} and \ref{QR} we respectively
depict $c_s^2$ and $Q_{\cal R}$, arising from the numerical elaboration
of the full system.

From the above analysis we can see that our model is stable under both
ghost and gradient instabilities. However, it generally generates a (deep)
blue tilted power spectrum, which cannot be consistent with
observations. Note that in \cite{Qiu:2011cy} this was shown for
$w=1/3$, therefore in our present case where $w>1$ the blue tilt of the
power spectrum is even stronger. This implies that the anisotropy-free
requirement $w>1$ and the scale-invariant perturbation generation  cannot
be obtained simultaneously in the current case, as was already mentioned
in the Introduction.
For this reason, we should introduce an additional mechanism in order to be
able to generate the nearly invariant perturbation spectrum, without
spoiling the solution to the anisotropy problem. This can be performed by
the curvaton mechanism, as we will present in the next section.

\section{The curvaton mechanism and the Scale-invariant spectrum}
\label{curvatonmechanism}

In the previous section we showed that the Galileon scenario under an
Ekpyrotic-like potential can exhibit a bouncing solution naturally. In the
contracting phase the corresponding total EoS of the universe satisfies
$w>1$ in order for the evolution to be free of the anisotropy problem
discussed in section \ref{anisotropyproblem}. However, under this
requirement the corresponding perturbations, although free of ghost and
gradient instabilities, cannot give rise to a nearly scale-invariant power
spectrum, as it is required by observational data
\cite{Bennett:2012fp}.

In \cite{Qiu:2011cy} it was shown that this problem can be solved by
introducing a curvaton field $\sigma$ coupled to the Galileon $\phi$. As it
is usual for curvaton fields \cite{Lyth:2001nq}, $\sigma$ does not affect
the bouncing background behavior, but it can lead to a power spectrum in
agreement with observations. In particular, through a specific
Galileon-curvaton coupling, the curvaton field lies effectively in a
``fake'' de-Sitter expansion or matter-like contraction, and thus it can
generate a nearly scale-invariant power spectrum. This is the idea behind
the ``conformal'' mechanism \cite{Hinterbichler:2011qk} and in this section
we investigate the necessary form of the coupling functions.


Let us consider the curvaton action as
\be
\label{action_curvaton}
{\cal S}_{\sigma}=\frac{1}{2}\int d^4x
\sqrt{-g}[-{\cal F}(\phi)(\partial\sigma)^2-2{\cal G}(\phi)W(\sigma)]~,
\ee
allowing for the most general form of coupling between $\sigma$ and $\phi$. The functions ${\cal F(\phi)}$ and ${\cal G(\phi)}$ depend on $\phi$, while $W(\sigma)$ is the potential for $\sigma$. Variation of action (\ref{action_curvaton}) with respect to $\sigma$ gives its background evolution equation as:
\be
\label{eomsigma}
\ddot\sigma_0+\frac{(a^3{\cal F})^\cdot}{a^3{\cal F}}\dot\sigma_0+\frac{{\cal G}}{{\cal F}}W_{\sigma_0}=0~,
\ee
where $\sigma_0$ is the background value of $\sigma$ and $W_{\sigma_0}$
corresponds to $\partial W/\partial\sigma|_{\sigma_0}$. In the following,
and up to the end of this section, we omit the subscript ``0'', denoting
the background by a simple $\sigma$, unless explicitly mentioned.
Additionally, the energy density and pressure of $\sigma$ can be
respectively written as:
\be
\label{rhosigma}
\rho_\sigma=\frac{1}{2}{\cal F}\dot\sigma^2+{\cal
G}W~,~P_\sigma=\frac{1}{2}{\cal F}\dot\sigma^2-{\cal G}W~.
\ee

We now proceed to the examination of the perturbations generated from the
curvaton field. Perturbing it by $\delta\sigma$ and defining $u\equiv
a\sqrt{{\cal F}}\delta\sigma$, we can extract the corresponding
perturbation equation as a second-order differential equation:
\be
u''+\left(k^2+a^2\frac{{\cal G}}{{\cal
F}}W_{\sigma\sigma}-\frac{z''}{z}\right)u=0~,
\ee
where the prime denotes derivative with respect to the conformal time $\eta$, $W_{\sigma\sigma}\equiv\partial^2W(\sigma)/(\partial\sigma)^2$ is the second derivative of the potential with respect to $\sigma$, and we have defined $z\equiv a\sqrt{\cal F}$. As usual, the power spectrum generated by $\delta\sigma$ is defined as:
\be
\label{sigmaspectrum}
{\cal P}_{\delta\sigma}=\frac{k^3}{2\pi^2}\Big|\frac{u}{z}\Big|^2~.
\ee
Therefore, we deduce that the condition for obtaining a scale-invariant power spectrum of $\delta\sigma$ is:
\be
\label{curvcon}
\frac{a^2{\cal G}W_{\sigma\sigma}}{\cal
F}-\frac{z''}{z}\simeq-\frac{2}{|\eta_\ast-\eta|^2}~.
\ee

There are several ways to satisfy the condition (\ref{curvcon}). The
simplest one is to set $W(\sigma)=0$, in which the first term in the above
condition disappears, and thus we need just to suitably choose ${\cal F}$
in order to obtain $z''/z\simeq2|\eta_\ast-\eta|^{-2}$. Alternatively we
can incorporate the
effects of both the kinetic and the potential terms of $\sigma$. In the following subsections we consider these cases separately.

\subsection{$W(\sigma)=0$}
\label{Wzero}

Under $W(\sigma)=0$, the condition (\ref{curvcon}) of obtaining scale-invariant power spectrum becomes
\be
\label{z}
z\propto |\eta_\ast-\eta|^2\,\,\,{\text{or}}\,\,\,|\eta_\ast-\eta|^{-1}~.
\ee

For the case of $z\propto |\eta_\ast-\eta|^2$ we obtain
\be
\delta\sigma=\frac{u}{z}\sim k^\frac{3}{2}~,~~~k^{-\frac{3}{2}}|\eta_\ast-\eta|^{-3}~,
\ee
the latter of which dominates over the former. Therefore, using expression
(\ref{sigmaspectrum}) we can obtain the power spectrum as
\be
{\cal P}_{\delta\sigma}\sim k^0 |\eta_\ast-\eta|^{-6}~.
\ee
As we observe, the spectrum is indeed scale-invariant but it has an increasing amplitude.

On the other hand, for the case of $z\propto|\eta_\ast-\eta|^{-1}$ we acquire
\be
\delta\sigma=\frac{u}{z}\sim k^{-\frac{3}{2}}~,~~~k^\frac{3}{2}|\eta_\ast-\eta|^{3}~,
\ee
the former of which dominates over the latter. Therefore, using
(\ref{sigmaspectrum}) we can obtain the power spectrum as
\be
{\cal P}_{\delta\sigma}\sim k^0.
\ee
In this case the spectrum is scale-invariant and moreover it is conserved on super-horizon scales.

The absence of $W(\sigma)$ leads to an absence of ${\cal G}(\phi)$ too.
Thus, we only need to suitably determine the form of ${\cal F}(\phi)$
according to the condition (\ref{z}). Note that far before the bounce we
have already assumed the scale-factor ansatz (\ref{scalea}), and therefore
 (\ref{z}) requires just
\be
\label{scalef}
{\cal F}\propto
|\eta_\ast-\eta|^\frac{2(2-3p)}{1-p}\,\,\, {\text{or}}
\,\,\,|\eta_\ast-\eta|^\frac{2}{p-1}~.
\ee
Furthermore, from the ansatz solution (\ref{ansatz}) for $\phi$ and the above expressions we deduce that the suitable choice of the form of ${\cal F}$ in terms of $\phi$ might be
\be
\label{f}
{\cal F}\propto e^{2(3-c^2)\phi/c}\,\,\, {\text{or}}\,\,\,e^{c\phi}~,
\ee
where we have made use of the relation $p=2/c^2$ (note that since in contracting phase $w\gg 1$, we have $0<p\ll 1$).

We close this subsection by examining the backreaction of $\sigma$ field on
the background evolution, since although the curvaton is necessary for the
correct perturbation generation we would not desire it to spoil the
background bouncing behavior itself. In the contraction region where
$t<t_\ast$, the background energy density of the system scales as
$\rho_{t<t_\ast}\sim (t_\ast-t)^{-2}$, as it was found in
(\ref{energybg}). On the other hand, the evolution equation of the curvaton
field (\ref{eomsigma}) gives $\dot\sigma\sim a^{-3}{\cal F}^{-1}$, and thus
its energy density in (\ref{rhosigma}) becomes $\rho_\sigma\simeq{\cal
F}\dot\sigma^2/2 \sim a^{-6}{\cal F}^{-1}$. Since we know the
time-dependence of the scale factor from (\ref{scalea}) and the
time-dependence of ${\cal F}$ from (\ref{scalef}), we straightforwardly
deduce that for the solution branch where $z\propto|\eta_\ast-\eta|^2$ we
obtain $\rho_\sigma\sim(t_\ast-t)^{-4}$,
while for the solution branch where $z\propto|\eta_\ast-\eta|^{-1}$ we acquire $\rho_\sigma\sim(t_\ast-t)^{2(1-3p)}\sim (t_\ast-t)^2$ (in the last step we used that $p\ll 1$). Therefore, our analysis indicates that in the solution branch where $z\propto|\eta_\ast-\eta|^{2}$ (with ${\cal F}\propto e^{2(3-c^2)\phi/c})$, one has to suitably tune the initial conditions in order for the curvaton not to destroy the background bouncing behavior. However, in the second solution branch where $z\propto|\eta_\ast-\eta|^{-1}$ (with ${\cal F}\propto e^{c\phi})$, the energy density of the curvaton field grows slower than that of the background, and thus the background bouncing evolution is not altered by the backreaction of the curvaton.

\subsection{$W(\sigma)\neq0$}
\label{Wnonzero}

We now examine the case where the curvaton potential is non-zero, in order to investigate its effect on the perturbation generation. Without loss of generality and for simplicity we assume that ${\cal F}$ is approximately a constant (we set ${\cal F}=1$), although extension to general ${\cal F}$ is straightforward.

In order to see what condition (\ref{curvcon}) gives in this case, we
recall that at the early stage of the bouncing phase, where the
perturbation $\delta\sigma$ is generated, the scale factor evolves
according to (\ref{scalea}), and therefore since $z\equiv a\sqrt{{\cal
F}}$ we obtain
\be
\label{zWnonzero}
\frac{z''}{z}=\frac{a''}{a}\simeq\frac{p}{1-p}\left(\frac{p}{1-p}
-1\right)\frac{1}{\left|\eta-\eta_{\ast}\right|^{2}}~.
\ee
Furthermore, introducing $a_\ast=a(\eta_\ast)$ we can write
\be
\label{a2Wnonzero}
a^{2}\frac{{\cal G}}{{\cal F}}W_{,\sigma\sigma}=a_\ast^2{\cal
G}W_{,\sigma\sigma}\left|\eta_{\ast}-\eta\right|^{\frac{2p}{1-p}}~.
\ee
Inserting equations (\ref{zWnonzero}) and (\ref{a2Wnonzero}) into
(\ref{curvcon}) we deduce that the condition for obtaining a
scale-invariant power spectrum of $\delta\sigma$ becomes
\be
\label{condWnonzero}
a_\ast^2{\cal
G}W_{\sigma\sigma}=\frac{3p-2}{(1-p)^2}\left|\eta_{\ast}-\eta\right|^{
-\frac{2}{1-p}}~.
\ee

As a specific example we consider the well-studied case of a quadratic
potential, namely $W(\sigma)=m_\sigma^2\sigma^2/2$, in which case
$W_{\sigma\sigma}=const$. Hence, condition (\ref{condWnonzero}), using
also the $\phi$-evolution from (\ref{ansatz}), gives
\be
\label{GWnonzero}
{\cal G}\propto e^{c\phi}~.
\ee
Therefore, we extract that the field perturbation $\delta\sigma$ scales as:
\be
\delta\sigma=\frac{u}{z}\sim
k^{-\frac{3}{2}}|\eta_\ast-\eta|^{\frac{1}{p-1}}~,~~~k^\frac{3}{2}
|\eta_\ast-\eta|^{\frac{3p-2}{p-1}}~.
\ee
We mention that since we assume ${\cal F}=1$ (that is $z=a$) the ``fake'' effect is absent, and thus the dominating mode of the perturbations is always the growing mode.

We close this subsection by examining the backreaction of $\sigma$ field on the background evolution, since we would not want the curvaton to spoil the background bouncing behavior itself. The $\sigma$-evolution equation (\ref{eomsigma}) becomes
\be
\label{sigmaeqWnonzero}
\ddot\sigma+3H\dot\sigma+m_\sigma^2{\cal G}\sigma=0~.
\ee
Since according to relations (\ref{GWnonzero}) and  (\ref{ansatz}) we have
${\cal G}\sim e^{c\phi}\sim(t_\ast-t)^{-2}$, we can write ${\cal G}={\cal
G}_0(t_\ast-t)^{-2}$ with ${\cal G}_0$ being an arbitrary constant.
Therefore, equation (\ref{sigmaeqWnonzero}) accepts the solution
\be
\label{solsigma}
\sigma\sim(t_\ast-t)^{\frac{1}{2}[1-3p\pm\sqrt{(1-3p)^2-4m_\sigma^2{\cal
G}_0}]}~.
\ee
Finally, substituting it into (\ref{rhosigma}) for the curvaton energy
density we obtain
\be
\label{curvenergWnonzero}
\rho_\sigma\sim(t_\ast-t)^{[1-3p\pm\sqrt{(1-3p)^2-4m_\sigma^2{\cal G}_0}]-2}~.
\ee

Comparing the background energy-density evolution (\ref{energybg}) with the curvaton energy-density evolution (\ref{curvenergWnonzero}) we deduce that the requirement for the latter to grow slower than the former is to have $1-3p\pm\sqrt{(1-3p)^2-4m_\sigma^2{\cal G}_0}>0$. However, since in order to solve the anisotropy problem we focus on $w>1$ (or equivalently $p<1/3$), then provided ${\cal G}_0>0$ the above requirement is always satisfied. Therefore, in the scenario at hand the energy density of the curvaton field will never dominate over the background evolution, that is the background bouncing behavior will not
be destroyed by its backreaction.

We close this section with a comment on the preservation of the scale
invariance across and after the bounce, which is in general a crucial
question  in bouncing scenarios, and on the matching conditions we impose. In
 the above analysis we required
$\delta\sigma$ and $\delta\sigma^\prime$ to be continuous across the bounce,
which is consistent with Deruelle-Mukhanov matching conditions
\cite{Deruelle:1995kd}. Considering the expanding phase, since it usually contains two
modes, namely the constant and the growing/decaying one, the perturbation
spectrum may or may not get altered
depending on whether mode-mixing is realized or not, or depending on
whether the varying modes in contracting/expanding phase are
growing/decaying, which is determined by the background. 
For instance, in the above
case where  ${\cal F}(\phi)\sim e^{c\phi}$ and $W(\sigma)=0$, the contracting
modes are decaying, thus by using the aforementioned matching conditions
scale-invariance will be maintained if the expanding mode is growing, which
requires the background EoS (or the effective EoS, if there is a ``faking'')
to be no larger than 1, and therefore we may easily preserve the
scale-invariance in the expanding phase by slightly constraining the
background evolution. A more detailed discussion on these will be taken
on in a following-up paper. Similar results can be found in
\cite{Cai:2012va}.

\section{Reconstructing the exact solution around the bounce}
\label{Reconstructing}

In the previous sections we constructed the Galileon bounce  scenario free of
the anisotropy problem, in which we added the curvaton field in order to
obtain a scale-invariant power spectrum of primordial perturbations generated
in the contracting phase. Additionally, we showed that under soft
requirements on the choice of ${\cal F}(\phi)$ or ${\cal G}(\phi)$, the
backreaction of the curvaton field will not alter the background bouncing
evolution.

However, after the bounce the effect of the curvaton field can  be
significant, and in particular it can regularize the universe evolution in
order not to result to a Big-Rip. In order to examine what classes of
coupling functions can provide this overall behavior, in this section we
semi-analytically reconstruct them following the ``inverse'' procedure
\cite{Cai:2009in}, in which we impose as input the desired bouncing scale
factor, reconstructing suitably the various function in order to correspond
to a consistent and exact solution of the full system of equations.

The complete action of the Galileon-curvaton system, consisted of both (\ref{action}) and (\ref{action_curvaton}), can be written as:
\bea
\label{fullaction}
&&{\cal
S}_{total}=\int
d^4x\sqrt{-g}\left[\frac{1}{2}R-\frac{1}{2}
\nabla_\mu\phi\nabla^\mu\phi-V(\phi)\right.\ \ \ \   \   \ \
\nonumber\\
&&\left.\ \ \ \   +\frac{g}{2}
\nabla_\mu\phi\nabla^\mu\phi
\Box\phi-{\cal F}(\phi)(\partial\sigma)^2-2{\cal
G}(\phi)W(\sigma)\right].
\eea
Note that although matter and radiation could be included
straightforwardly, in the above action we have  neglected them in order to
examine the pure effects of the Galileon-curvaton evolution. Thus, the
cosmological equations in the FRW metric are the first Friedmann equation:
\be
\label{FR1}
3H^2=\frac{\dot{\phi}^2}{2}+V(\phi)+3gH\dot{\phi}^3+\frac{1}{2}{\cal
F}(\phi)\dot{\sigma}^2+{\cal G}(\phi)W(\sigma)~,
\ee
and the evolution equations for the two fields, namely
\be
\label{sigmaevol}
\ddot{\sigma}+\dot{\sigma}\left[3H+\frac{\partial{\cal
F}(\phi)}{\partial\phi}\frac{\dot{\phi}}{{\cal F}(\phi)}\right]+\frac{{\cal
G}(\phi)}{{\cal F}(\phi)}\frac{\partial W(\sigma)}{\partial\sigma}=0~
\ee
and
\bea
\label{phievol}
 &&\ddot{\phi}
\left[1+6gH\dot{\phi}+\frac{3}{2}g^2\dot{\phi}^4\right]-\frac{1}{2}\frac{
\partial{\cal F}(\phi)}{\partial\phi}\dot{\sigma}^2+\frac{\partial{\cal
G}(\phi)}{\partial\phi}W(\sigma)\nonumber\\
&&+\frac{3}{2}\dot{\phi}\left\{2H+g\dot{\phi}\left[6H^2-\dot{\phi}^2-{\cal
F}(\phi)\dot{\sigma}^2-3gH\dot{\phi}^3
\right]\right\}\nonumber\\&&+\frac{\partial V(\phi)}{\partial\phi}=0~,
\eea
respectively.

One could solve the above equations fully numerically, imposing specific
ansantzes and initial conditions, however doing so he does not have
control of what functions-ansantzes, parameter choices and initial
conditions, lead to bouncing solutions. That is why the aforementioned,
semi-analytical, ``inverse'' procedure, where the scale factor is imposed
{\it a priori}, is better and more appropriate for the analysis of this
section, allowing for a systematic control on the conditions of the bounce
realization. We mention that since there are more unknown functions than
equations, one can in general always reconstruct the desired evolution.

Let us impose a desired bouncing scale factor $a(t)$ as an input, by which
$H(t)$ is also known. Furthermore, we consider $V(\phi)$ as usual, and we
impose $\phi(t)$ at will too. Thus, only three free functions remain,
namely ${\cal F}(\phi)$, ${\cal G}(\phi)$ and $W(\sigma)$ which must be
derived by the equations, along with the solution for $\sigma(t)$. Since
there are three independent cosmological equations, namely equations
(\ref{FR1}), (\ref{sigmaevol}) and (\ref{phievol}), we must also impose by
hand one more of the above four functions. We prefer to set $W(\sigma)=0$,
since this was one (simpler) case that was approximately analyzed in
subsection \ref{Wzero} (however one could easily consider other $W(\sigma)$
forms too). Such a choice simplifies things since the function ${\cal
G}(\phi)$ also disappears from the equations, and therefore the
cosmological equations (\ref{FR1})-(\ref{phievol}) are considered as
differential equations for $\sigma(t)$ and ${\cal F}(t)$. Thus, after
obtaining the solution, and since we know $\phi(t)$, we can reconstruct
${\cal F}(\phi)$.

Equation (\ref{FR1}) can be algebraically solved in order to obtain
$\dot{\sigma}^2$ as
\bea \dot{\sigma}^2(t)&=&\frac{1}{{\cal
F}(t)}\left[6H(t)^2-2V(\phi(t))-\dot\phi^2(t)\right.\nonumber\\
\label{auxil4}
&&\left.\ \ \ \ \ \ \ \  -6gH(t)\dot\phi^3(t)\right]~.
\eea
Substituting this into (\ref{sigmaevol}) gives a simple first
order differential equation for ${\cal F}(t)$ of the form
\be\label{auxil5}
h(t,{\cal F}(t),\dot{\cal F}(t))=0~,
\ee
which can be easily solved. Thus, from the solution of ${\cal F}(t)$ and the known $\phi(t)$ we can reconstruct ${\cal F}(\phi)$.

We mention here that the above procedure holds for every input functions,
with the only requirement being the obtained $\dot{\sigma}^2$ from relation
(\ref{auxil4}) to be positive, otherwise there is no solution that can
correspond to the input functions. With the above semi-analytical procedure
one has full control on how to choose the model parameters and the initial
conditions in order to get a positive $\dot{\sigma}^2$ in
(\ref{auxil4}). On the other hand, if one tries to solve fully numerically
the three equations (\ref{FR1})-(\ref{phievol}) simultaneously, it is very
hard to determine the model parameters and the initial conditions in order
to get a consistent bouncing solution.

In order to apply explicitly the above reconstructing procedure, without loss of generality we choose a bouncing scale factor of the form
\be
\label{at}
a(t)=a_B\left(1+\frac{3}{2}\omega t^{2}\right) ^{1/3}~,
\ee
where $a_B$ is the scale factor at the bouncing point and $\omega$ is a
positive parameter which describes how fast the bounce takes place. The
above ansatz presents the bouncing behavior, where $t$ varies in the
bounce region, that is between the times $t_{B-}$ and $t_{B+}$, with
$t=t_B=0$ the bouncing point, however one could use at will any other
bouncing ansatz. Straightforwardly we find
\be
H(t)=\frac{\omega t}{(1+3\omega t^{2}/2)}~,
\ee
and thus the universe is free of a Big-Rip after the bounce.

 \begin{figure}[ht]
\includegraphics[scale=0.32]{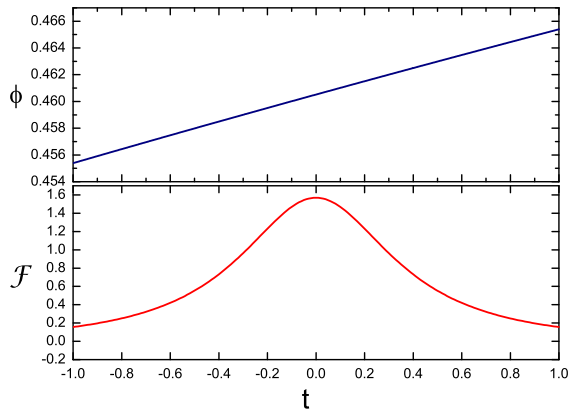}
\caption{(Colored online) {\it{The solution for the coupling function
${\cal F}(t)$ and the imposed Galileon field $\phi(t)$, under the imposed
bouncing ansatz (\ref{at}). We choose the parameters as $g=1$,
$V_0=1/12$, $c=2\sqrt{3}$, $\omega=1$, $\phi_I=0.1$, $t_I=-100$, and
$t_{B\pm}=\pm1$.}}}
\label{phiFtt}
\end{figure}
\begin{figure}[ht]
\includegraphics[scale=0.32]{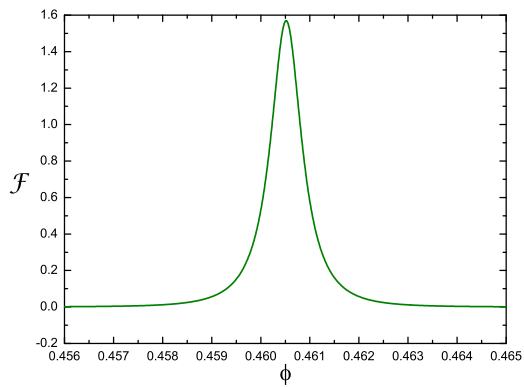}
\caption{(Colored online) {\it{The reconstructed  coupling function
${\cal F}(\phi)$ under the imposed
bouncing   ansatz  (\ref{at}), using Fig. \ref{phiFtt}. We choose
the parameters    as $g=1$, $V_0=1/12$, $c=2\sqrt{3}$, $\omega=1$,
$\phi_I=0.1$, $t_I=-100$, and $t_{B\pm}=\pm1$.}}}
\label{Fphi}
\end{figure}
For the field $\phi$ and the potential $V(\phi)$, enlightened by the
analysis of subsection \ref{galileonbackground} and without loss of
generality, respectively we assume
\be
\label{phisample}
\phi(t)=\phi_I\ln(t-t_I)~
\ee
and
\be
V(\phi)=-V_0e^{c\phi}~,
\ee
 while as we mentioned we set $W(\sigma)=0$.

\begin{figure}[ht]
\includegraphics[scale=0.32]{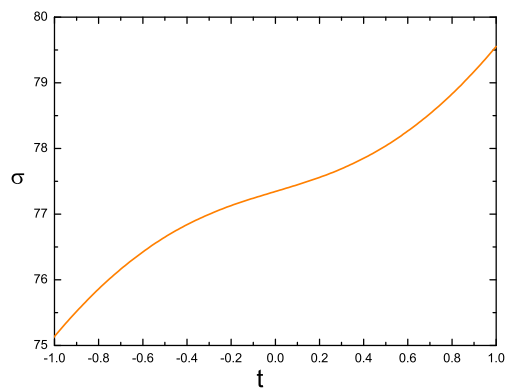}
\caption{(Colored online) {\it{The solution for the curvaton field
$\sigma(t)$, under the imposed
bouncing ansatz  (\ref{at}). We choose the
parameters   as $g=1$, $V_0=1/12$, $c=2\sqrt{3}$, $\omega=1$, $\phi_I=0.1$,
$t_I=-100$, and $t_{B\pm}=\pm1$.}}}
\label{sigma}
\end{figure}

We follow the procedure described above, and for the model parameters we
choose $g=1$, $V_0=1/12$, $c=2\sqrt{3}$, $\omega=1$, $\phi_I=0.1$,
$t_I=-100$, and $t_{B\pm}=\pm1$. In Fig. \ref{phiFtt} we depict the
solution for ${\cal F}(t)$ and also the known $\phi(t)$ from
(\ref{phisample}), and in Fig. \ref{Fphi} we present the corresponding
reconstructed ${\cal F}(\phi)$. Finally, for completeness in Fig.
\ref{sigma} we show the solution for $\sigma(t)$.

We close this section by mentioning that in principle one could think of other reconstructing procedures, for instance setting ${\cal F}(\phi)$ and reconstruct ${\cal G}(\phi)$ and $W(\sigma)$, or even setting $\phi(t)$ and $\sigma(t)$ and reconstruct ${\cal F}(\phi)$, ${\cal G}(\phi)$ and $W(\sigma)$. So there can actually be many possibilities to realize the non-singular bounce.

\section{Conclusions}
\label{Conclusions}

Bounce cosmology is an interesting paradigm since it alleviates the
Big-Bang singularity problem. Additionally, it can solve the
Big-Bang problems, and nearly scale-invariant primordial
perturbations can be incorporated too. These features make bouncing
cosmologies successful alternatives to inflation. However, there are many
detailed issues that should be carefully addressed during the establishment
of bouncing cosmology, and in the present work we tried to confront some of
them.

First of all, the NEC violation, which is required for the bounce
realization, may bring ghost degrees of freedom. In the above analysis we
were based on the Galileon scenario, which is a higher-derivative
construction free of ghosts, and thus we obtained a bouncing evolution free
of ghost and gradient instabilities.

However, there is a second problem that may disturb the bounce
construction, namely that in the contracting phase even a tiny anisotropic
fluctuation from the totally isotropic FRW geometry will be radically
enhanced and destroy completely the FRW evolution. The solution of this
``anisotropy problem'' requires the total EoS of the universe to lie in
the regime $w>1$ in the contracting phase. Thus, starting from
\cite{Qiu:2011cy} where the anisotropy problem was present, in this work we
were able to solve it and obtain $w>1$ by considering an Ekpyrotic-like
potential with negative value. This is one of the main contributions of the
present article.

The above solution of the anisotropy problem through a large EoS has an
undesired effect, namely it spoils the generation of a nearly
scale-invariant power spectrum, that a scalar with smaller EoS can bring
through adiabatic perturbation. Therefore, in order to still be able to
produce a power spectrum in agreement with observations, we additionally
introduced in the scenario a second, curvaton field, coupled to the
Galileon one, which can indeed generate the desired perturbations in an
isocurvature way. In our analysis we presented this mechanism in general,
and we analyzed explicitly two specific examples where nearly
scale-invariant perturbations are generated. Finally, we examined the
conditions under which the curvaton field does not cause a significant
backreaction on the background bouncing behavior caused by the Galileon
field.

Furthermore, the curvaton field, apart from the generation of the desired
perturbations, has another important role, namely after the bounce it can
regularize the background evolution in order to avoid a Big-Rip
singularity, which is caused by the Galileon field itself. In particular,
although the curvaton backreaction is  not significant at the background
level before the bounce, during and after the bounce it becomes important
and changes the background evolution. In order to see this effect
we performed a semi-analytically ``inverse'' analysis, reconstructing
suitably the desired bouncing evolution of the scale factor, which is free
of a Big-Rip   without any fine-tuning. We mention here that since
the region where the scale-invariant spectrum is generated lies in
the contracting phase, while the region where the Big-Bang is avoided is
around and after the bounce point, the conditions on the functions
that generate    scale-invariance perturbations should hold in the
contracting phase while those for the Big-Bang avoidance  should hold around
and after the
bounce. Therefore, one can always match the required function form of the
contracting regime with the required form of the bounce regime, to obtain
both perturbation scale invariance and  Big-Bang avoidance, although not
always analytically.

We close this work by mentioning that there could be other
possibilities to avoid the Big-Rip singularity. For instance, an
alternative evolution after the bounce would be to assume that the Galileon
and curvaton fields decay to standard model particles
\cite{Langlois:2013dh}. In this case the decaying Galileon energy density
cannot trigger the Big Rip anymore, and additionally it can produce the
matter content of the universe. Such a detailed analysis of the post-bounce
evolution, and its relation to the subsequent thermal history of the
universe, lies beyond the scope of the present work and it is left for a
future investigation.

{\bf Note added:} After completing our manuscript, we came to know that
studies of bounce cosmology aiming to the same issue has been done in
\cite{Osipov:2013ssa}, in which similar results are obtained for different
(conformal) Galileon models but with the same (Ekpyrotic-like) potential.

\begin{acknowledgments}
T.Q. thanks Robert Brandenberger for his useful comments. The work of T.Q.
is funded in part by the National Science Council of R.O.C. under Grant No.
NSC99-2112-M-033-005-MY3 and No. NSC99-2811-M-033-008 and by the National
Center for Theoretical Sciences. X.G. was supported by ANR (Agence Nationale
de la Recherche) grant ``STR-COSMO" ANR-09-BLAN-0157-01. The research of
E.N.S. is implemented within the framework of the Action ``Supporting
Postdoctoral Researchers'' of the Operational Program ``Education and
Lifelong Learning'' (Actions Beneficiary: General Secretariat for Research
and Technology), and is co-financed by the European Social Fund (ESF) and
the Greek State.
\end{acknowledgments}


\end{document}